\documentclass[reprint,
amsmath,
amssymb,
aps,
prd,
twocolumn,
superscriptaddress,
nofootinbib,
floatfix,
preprintnumbers,
showpacs,
]{revtex4-1}

\usepackage{graphicx,amssymb,amsmath,subfigure}
\usepackage{dcolumn}
\usepackage{bm}
\usepackage{natbib}
\usepackage{hyperref}
\usepackage{ulem}
\usepackage{color}

\begin{document}

\preprint{DESY 16-139}

\title{Non-thermal Axion Dark Radiation and Constraints}

\author{Anupam Mazumdar}
\email{a.mazumdar@lancaster.ac.uk}
\affiliation{Consortium for Fundamental Physics, Physics Department, Lancaster University}
\affiliation{Kapteyn Astronomical Institute, University of Groningen, 9700 AV Groningen, The Netherlands}

\author{Saleh Qutub}
\email{s.qutub@lancaster.ac.uk}
\affiliation{Consortium for Fundamental Physics, Physics Department, Lancaster University}
\affiliation{Department of Astronomy, King Abdulaziz University, Jeddah 21589, Saudi Arabia}

\author{Ken'ichi Saikawa}
\email{kenichi.saikawa@desy.de}
\affiliation{Deutsches Elektronen-Synchrotron DESY, Notkestrasse 85, 22607 Hamburg, Germany}

\date{\today}

\begin{abstract}
The Peccei-Quinn  mechanism presents a neat solution to the strong CP problem. As a by-product, it provides an ideal dark matter candidate, ``the axion", albeit with a tiny mass. Axions therefore can act as dark radiation if excited with large momenta after the end of inflation. Nevertheless, the recent measurement of 
relativistic degrees of freedom from cosmic microwave background radiation strictly constrains the abundance of such extra relativistic species. We show that ultra-relativistic axions can be abundantly produced if the Peccei-Quinn field was initially displaced from the minimum of the potential. This in lieu places an interesting constraint on the axion dark matter window with large decay constant which is expected to be probed by future experiments. Moreover, an upper bound on the reheating temperature can be placed, which further constrains the thermal history of our Universe.

\end{abstract}

\pacs{14.80.Va, 95.35.+d, 98.80.Cq}

\maketitle

\section{Introduction} \label{sec:intro}

The strong CP problem is one of the outstanding puzzle of particle physics today. It is a well-known fact that quantum chromodynamics (QCD) allows a CP violating term of the form $ \theta (g_s^2/32 \pi^2) G^{b  \mu \nu} \widetilde{G}^b_{\mu \nu}$, where
$\theta$ is a constant parameter~\cite{'tHooft:1976up}. The stringent bound on the electric dipole moment of neutron implies that $\vert \theta \vert < 0.7 \times 10^{-11}$~\cite{Crewther:1979pi}. Such a small value for $\theta$ is quite unnatural. This problem can be elegantly solved by the Peccei-Quinn (PQ) mechanism in which a global U(1)$_{\rm PQ}$ symmetry with a chiral anomaly is introduced and the CP violating $\theta$-term can be dynamically relaxed to zero~\cite{Peccei:1977ur}. The corresponding Goldstone boson, the axion~\cite{Weinberg:1977ma} remains massless at the classical level, but acquires a periodic potential and consequently a mass inversely proportional to the PQ symmetry breaking scale, $f_{\rm PQ}$, due to non-perturbative QCD effect~\cite{'tHooft:1976up}.

Even though the original PQ proposal with $f_{\rm PQ}$ at the electroweak (EW) scale was soon ruled out by several experiments~\cite{Cheng:1987gp}, other variants of the PQ mechanism 
circumvent the problem by creating a hierarchy between the PQ breaking scale and the EW one via the introduction of a new Standard Model (SM) complex singlet field whose vacuum expectation value (VEV) breaks the PQ symmetry~\cite{Kim:1979if,Dine:1981rt}. The scale of PQ symmetry breaking is subjected to several observational  constraints. For instance, the observation of the supernova SN1987A, white dwarfs and the globular clusters set a lower bound of $(2 \textendash 4) \times 10^{8}$~GeV on the scale of PQ symmetry breaking
~\footnote{More precisely, on the axion decay constant, $f_a = f_{\rm PQ}/N_{\rm DW}$, where $N_{\rm DW}$ is the number of domain walls; $N_{\rm DW} \geq 1$ for KSVZ models~\cite{Kim:1979if}, and $N_{\rm DW}  = 6$ for DFSZ models~\cite{Dine:1981rt}. We elaborate on these models in Appendix-A and B. } (see e.g.~\cite{Raffelt:1996wa,Raffelt:2006cw} and references therein).
With such a high PQ symmetry breaking scale, axion can be a good dark matter (DM) candidate~\cite{Preskill:1982cy}, whose couplings to other fields are suppressed by powers of $f_{\rm PQ}$.  In fact, the energy stored in the coherent oscillations of axions today can make the entirety of the observed DM abundance for
$f_{\rm PQ} \sim 7 \times 10^{11} \ {\rm GeV} \ N_{\rm DW} \langle \theta_i^2 \rangle^{-0.84} $~\cite{Turner:1985si} where $ \langle \theta_{i}^2 \rangle$
is the axion misalignment angle at beginning of the axion oscillation phase.

On the other hand, many puzzles of early Universe cosmology can be solved by an early epoch of accelerated expansion, ``inflation'' (for a review, see~\cite{Mazumdar:2010sa}). Inflation is also responsible for seeding the primordial  perturbations for cosmic microwave background (CMB) and large scale structures. During inflation, if there
exists any light field, such as moduli, whose mass is below ${\cal O}(\mathcal{H}_{\rm inf})$, they obtain vacuum induced quantum fluctuations 
 of ${\cal O}({\cal H}_{\rm inf}/2 \pi)$~\cite{Starobinsky:1986fx}, where $\mathcal{H}_{\rm inf}$ denotes the Hubble parameter during inflation. In this case, such 
 a light moduli can obtain large VEV, i.e. ${\cal O}(M_{\rm P})$, where $M_{\rm P} \simeq 2.43 \times 10^{18}$~GeV is the reduced Planck mass. Typically, the moduli behaves like a condensate within our Hubble patch~\cite{Enqvist:2003gh}, and begins its coherent oscillations when the Hubble expansion rate of the Universe drops to the mass of the moduli.

Similarly, if the PQ field is light compared to the Hubble expansion rate during inflation, then the PQ field can also be displaced from its minimum, which is 
determined by $f_{\rm PQ}$~\cite{Starobinsky:1986fx,Linde:1991km}, and consequently after the end of inflation the PQ field will start coherent oscillations when its 
mass would exceed the time dependent Hubble scale.
The PQ field can also be displaced away from $f_{\rm PQ}$ during inflation if it is coupled to the inflaton field~\cite{Linde:1991km}, and later starts oscillating once inflaton begins its own coherent oscllations around the minimum of its potential.
If the initial VEV of PQ field during inflation is displaced by $\gg f_{\rm PQ}$, the initial phase of oscillation takes place around the origin. This can lead to the restoration of the PQ symmetry and formation of dangerous topological
defects~\cite{Zel'dovich:1975,Sikivie:1982qv}~\footnote{For $N_{\rm DW} = 1$ (i.e. for KSVZ-like models with only one extra heavy quark species), these defects are unstable and decay to cold axions leading to an upper bound on the PQ breaking scale, $f_{\rm PQ} \lesssim (4.6 \textendash 7.2) \times 10^{10} \,{\rm GeV}  ( \Omega_a/\Omega_{\rm CDM} )^{0.84}$,
where $\Omega_a$ denotes the cold axions abundance and $\Omega_{\rm CDM}$ is the observed abundance of cold dark matter (CDM)~\cite{Hiramatsu:2012gg,Kawasaki:2014sqa}. On the other hand, when $N_{\rm DW} > 1$ (i.e. for DFSZ-like models or KSVZ-like models with several extra heavy quark species), the topological defects are stable and dominate the energy density of the Universe ruling out such scenario unless one fine-tunes a bias term that explicitly breaks the shift symmetry, and in this case $f_{\rm PQ}$ is constrained to be less than $\mathcal{O}(10^{10})\,{\rm GeV}$ in order to avoid the overproduction of axions~\cite{Kawasaki:2014sqa,Hiramatsu:2012sc}. }.

The non-thermal restoration of the PQ symmetry
can be avoided if the amplitude of the PQ field at the beginning of the oscillation phase is less than $\lesssim 10^4 f_{\rm PQ}$~\cite{Kawasaki:2013iha} or, if there is a coupling between 
the PQ field and the total energy density of the inflaton,
and the oscillation of the PQ field is driven by a higher order term in the potential~\cite{Harigaya:2015hha,Kearney:2016vqw}. Note that 
similar constraints would follow, if we had considered a moduli field instead of an inflaton field.

Once the amplitude of the PQ field drops below $f_{\rm PQ}$, the oscillation of the field continues around its minimum at $f_{\rm PQ}$. 
In such a case, there will be no non-perturbative production of QCD axion 
during the second phase of the oscillation~\cite{Mazumdar:2015pta}, 
but it can still lead to dangerous consequences from a perturbative decay of the PQ field.

In a wide range of parameter space, the coherent oscillation of the PQ field can decay dominantly into ultra-relativistic axions. If this decay occurs at sufficiently late times, the resultant axions will not thermalise with the plasma keeping their initial abundance and momenta. Such hot axions will act as
dark radiation, increasing the effective number of relativistic degrees of freedom (dof), i.e. $N_{\rm eff}$. The value of $N_{\rm eff}$ is constrained by the observation of CMB~\cite{Ade:2015xua}, allowing us to put constraints on the PQ parameter space.

The constraints on the extra relativistic species induced by heavy decaying particles are extensively discussed in the literature. These include the 
discussion in the context of the supersymmetric axion models~\cite{Kawasaki:2007mk}, and in the context of the heavy moduli decay in string cosmology~\cite{Cicoli:2012aq,Cicoli:2010ha}, where reheating SM degrees of freedom remains a challenge. 

Instead of considering these scenarios, where the mass of decaying particles
is generically controlled by the supersymmetry breaking effects, here we focus on the standard non-supersymmetric PQ mechanism, in which the decaying 
particle is identified as the radial component of the PQ field. The mass of the radial field is determined by the self coupling constant and the PQ symmetry breaking scale.
We show that an upper bound on the reheating temperature can be placed, which
is relevant to axion DM with a large decay constant.

This paper is structured as follows. In section~\ref{sec:setup}, we review the dynamics of the PQ symmetry breaking followed by a brief review of axion thermalisation and thermal production in section~\ref{sec:th-prod}. In section~\ref{sec:nonth-prod}, we discuss the non-thermal production of ultra-relativistic axions from the coherent oscillation of the radial component of the PQ field. We discuss the different constraints on the axion parameter space in section~\ref{sec:constraints}. Finally, we conclude our discussion in section~\ref{sec:conclusion}.

\section{Dynamics of the PQ symmetry breaking and the coherent oscillation of the PQ field} \label{sec:setup}

Let us now consider the dynamics of the PQ symmetry breaking during the evolution of a real scalar field $\phi$ in the background. The field $\phi$ could be inflaton or moduli as such. Our main focus in this paper is to understand the dynamics of PQ field and $\phi$ field after inflation.
The PQ symmetry breaking can be realized via the following Mexican hat potential for the PQ field denoted below by $S$.
\begin{equation} \label{eq:pot1}
V(\phi,S ) =
\lambda \left[ \vert S \vert^2 - \frac{f_{\rm PQ}^2}{2}  \right]^2 \!\! - g \, \phi^2 \vert S \vert^2 + U(\phi)
\, ,
\end{equation}
where $g, \lambda >0$~and  $f_{\rm PQ} = N_{\rm DW} f_a$ is the PQ breaking scale with $f_a$ and $N_{\rm DW}$ being the axion decay constant and the
domain walls number, respectively, 
and $U(\phi)$ is the potential of the scalar field $\phi$ which can be approximated by a quadratic one~\footnote{Here we consider a negative  coupling to $\phi$, i.e. $g>0$, in which case the effective PQ breaking scale, $f_{\rm PQ,eff} \equiv \sqrt{2} \langle \vert S \vert \rangle$, can be much larger than $f_{\rm PQ}$ causing the PQ field to oscillate once the slow-roll conditions break down, if $\phi$ was treated as an inflaton. This can also
ameliorate
the isocurvature bound on $f_{\rm PQ}$,
since large $f_{\rm PQ,eff}$ reduces the power spectrum of isocurvature perturbation along the angular direction given by ${\cal P}_{\rm Sc} = (4/\langle \theta_i^2 \rangle)(N_{\rm DW} {\cal H}_{\rm inf}/2 \pi f_{\rm PQ,eff})^2 (\Omega_a/\Omega_{\rm CDM})^2$~\cite{Harigaya:2015hha}, which is bounded to be $< 7.8 \times 10^{-11}$ from CMB data~\cite{Ade:2015xua}.
On the other hand, if $g<0$ and $\sqrt{g} \Phi_0/(\sqrt{\lambda} f_{\rm PQ}) \gtrsim {\cal O}(1)$, the PQ symmetry gets broken
after inflation
 leading to the formation of topological defects. However, the dynamics of the oscillations should not be different in either case.} 
\begin{equation} \label{eq:phi_potential}
U(\phi) = \frac{1}{2} m_\phi^2 \phi^2\,.
\end{equation}
A minor departure from a quadratic potential will not affect our discussion 
once $\phi$ starts oscillating around its minimum.
Note that coupling in Eq.~(\ref{eq:pot1}),  $g \phi^2 \vert S \vert^2$, will shift
 the minimum of the 
PQ field to:  $ \vert S \vert_{\rm m} = [ f_{\rm PQ}^2/2 +  (g/2 \lambda) \phi^2 ]^{1/2}$.

For convenience, let us write the PQ field in terms of polar coordinates, 
\begin{equation}
S= \frac{\sigma}{\sqrt{2}} e^{i \theta}. 
\end{equation}
The equations of motion for $\phi$ and the radial field $\sigma$ are then given by
\begin{eqnarray}
\label{eq:eomphi}
&&\Box \phi + 3 {\cal H} \dot \phi+ \partial_\phi U - g \sigma^2 \phi=0
\, ,
\\
\label{eq:eomsigma}
&&\Box \sigma + 3 {\cal H} \dot \sigma + \left[ \lambda(\sigma^2 - f_{\rm PQ}^2) - g \phi^2 \right] \sigma =0
\, ,
\end{eqnarray}
where dot denotes the derivative with respect to physical time $t$, and $\Box = \partial_t^2-\nabla^2/{\rm R}^2(t)$ with ${\rm R}(t)$ being the scale factor. In Eq.~(\ref{eq:eomsigma}), we have ignored an irrelevant coupling to the angular field $\theta$, $\sigma^2 \partial_\mu \theta \partial^\mu \theta$. Furthermore, we focus on the evolution of the zero-modes.
Separating the background part from Eqs.~(\ref{eq:eomphi}) and (\ref{eq:eomsigma}) by writing $\phi$ as $\bar{\phi}(t) + \delta \phi(t,{\bf x})$ and similarly for $\sigma$,  
we obtain
\begin{eqnarray}
\label{eq:eomphi0}
&&\ddot{\bar \phi} + 3 {\cal H} \dot{\bar \phi}+ [ m_\phi^2 - g \bar \sigma^2 ] \bar \phi=0
\, ,
\\
\label{eq:eomsigma0}
&&\ddot{\bar \sigma} + 3 {\cal H} \dot{\bar \sigma} + \left[ \lambda(\bar \sigma^2 - f_{\rm PQ}^2) - g \bar \phi^2 \right] \bar \sigma =0
\,,
\end{eqnarray}
where ${\cal H}={\rm \dot R(t)}/{\rm R(t)}$, and dot denotes time derivative with respect to physical time. Over-barred quantities are the background values. Now depending on the value of the parameters $g$ and $\lambda$, there will be two cases.

\subsection{Case I: $\sqrt{g} \Phi_0/(\sqrt{\lambda} f_{\rm PQ}) \ll 1$} \label{sub:case1}

In the limit when 
$\sqrt{g} \Phi_0/(\sqrt{\lambda} f_{\rm PQ}) \ll 1$,
where $\Phi_0$ denotes the amplitude of $\phi$ at the beginning of the oscillation phase after the end of inflation,
the coupling between the PQ field and $\phi$ can be ignored. 
In this case 
 the minimum of the potential along the $\sigma$ direction occurs at $f_{\rm PQ}$. 
Assuming it starts from a large value~\cite{Linde:2005ht, Kolb:1990vq}, the $\sigma$ field follows an attractor solution~\cite{Harigaya:2012up}
\begin{equation}
\sigma = \left[ 2 \lambda \int_{\phi}^{\phi_*} U_\phi^{-1} d\phi \right]^{-1/2}
\, .
\end{equation}
Clearly,
the effective PQ breaking scale, $f_{\rm PQ, eff}=  \langle \sigma \rangle $ can be much larger than $f_{\rm PQ}$ during the slow-roll phase. If $\phi$ is the inflaton, large $f_{\rm PQ, eff}$  ($\gg {\cal H}_{\rm inf}$) is actually desirable in order to suppress the isocurvature fluctuations along the angular direction.

We demand that the PQ field does not come to dominate the energy density of the Universe during the slow-roll phase, which constraints
\begin{equation}
\lambda \lesssim \frac{{\cal H}_{\rm inf}^2 M_P^2}{ \sigma_0^4} \label{eq:lambdamax}
\, .
\end{equation}
Here, $\sigma_0$ is the typical
value
of $\sigma$ during inflation.
Once the inflationary slow-roll conditions break down, the PQ field starts oscillating. Due to  large amplitude, $\sigma_0\gg f_{\rm PQ}$, the initial oscillations will take place around $\sigma = 0$. This can result in large quantum amplification of the fluctuations in the PQ field especially along the massless angular direction~\cite{Mazumdar:2015pta}, which may lead to the restoration of the PQ symmetry and consequently the formation of potentially dangerous topological defects unless $\sigma \lesssim 10^4 f_{\rm PQ}$~\cite{Kawasaki:2013iha}. Note that we shall adhere to this bound on $\sigma \lesssim 10^4 f_{\rm PQ}$ here.

\subsection{Case II:  $\sqrt{g} \Phi_0/(\sqrt{\lambda} f_{\rm PQ}) \gg 1$ } \label{sub:case2}

In this case the minimum of the PQ field gets shifted away from $f_{\rm PQ}$ via the coupling $ g \sigma^2 \phi^2$,
\begin{equation}
\sigma_{\rm m}   \simeq (g/ \lambda)^{1/2} \ \phi\,.
\end{equation}
If the PQ radial field is sufficiently heavy,
$ m_{\sigma, \rm eff} \simeq \vert \lambda(3 \sigma^2 - \sigma_{\rm m}^2) \vert^{1/2} \gtrsim {\cal H}_{\rm inf} $,
it will be sitting at its minimum,
$\sigma = \sigma_{\rm m} \gg f_{\rm PQ}$,
during the slow-roll phase. On the other hand, if $\sigma$ is light, $ m_{\sigma, \rm eff} \ll {\cal H}_{\rm inf} $, it can get displaced from $\sigma_{\rm m}$, i.e. $\sigma$ get shifted even further away from $f_{\rm PQ}$ due to inflaton vacuum induced quantum fluctuations. This result in an even larger initial amplitude of $\sigma$ once it start oscillating.
We follow Ref.~\cite{Harigaya:2015hha}, and take the amplitude of the PQ radial field at the beginning of the oscillation phase to be $\sigma_{\rm m}$.
Once
the PQ field starts oscillating around its minimum, the PQ field sooner or later stops tracking its minimum to oscillate around the origin 
with its own frequency. To see 
this,  let us define $ \xi = \bar \sigma /\sigma_{\rm m}$. Substituting it in Eq.~(\ref{eq:eomsigma0}), we obtain
\begin{equation}
\frac{d^2 \xi}{d (\ln t)^2} + F \frac{d \xi}{d (\ln t)}
+ ({\cal H} t)^2 \left[ \lambda  \frac{\sigma_{\rm m}^2}{{\cal H}^2} (\xi^2 - 1) + G \right] \xi = 0
\, ,
\end{equation}
with
\begin{equation}
F = {\cal H} t \left( \frac{2}{{\cal H} \sigma_{\rm m}} \frac{d \sigma_{\rm m}}{dt} + 3 \right) - 1
\simeq -1
\nonumber
\, ,
\end{equation}
and
\begin{equation}
G  = \frac{1}{\sigma_{\rm m} {\cal H}^2} \frac{d^2 \sigma_{\rm m}}{dt^2} + \frac{3}{\sigma_{\rm m} {\cal H}} \frac{d \sigma_{\rm m}}{dt}
\nonumber
\, ,
\end{equation}
where we substituted $\sigma_{\rm m} = \sqrt{(g/\lambda)\langle\phi^2\rangle}\propto t^{-1}$
with $\langle\phi^2\rangle$ being the time average of $\phi^2$ over its oscillation period.
Initially,
the PQ radial field is following its minimum $\sigma_{\rm m}$, i.e. $ \xi$ is roughly constant in time. Upon the breakdown of the slow-roll conditions, $\phi$ starts oscillating with frequency $m_\phi$ and amplitude decaying as ${\rm R}^{3/2}$. Due to its coupling to $\phi$ ($g \phi^2 \sigma^2 \propto {\rm R}^{-5}$), the PQ radial field may continue tracking $\sigma_{\rm m}$ for a while.
As the damping coefficient $F$ becomes negative once $\phi$ starts oscillating, the $\sigma$ tracking of its minimum is rendered unstable and the amplitude of $\xi$ will increase with time. Once the term $\lambda \sigma^4 \propto {\rm R}^{-4}$ takes over, the amplitude of $\xi$ will continue increasing as ${\rm R}^{1/2}$ and the oscillation of $\sigma$ will follow the solution~\cite{Greene:1997fu}
\begin{equation}
\sigma \sim \frac{\sigma_0}{{\rm R(\tau)}} \cos [ c \, \sqrt{\lambda} \, \sigma_0 (\tau - \tau_0) ]
\, ,
\end{equation}
where $c \simeq 0.8472$ is a constant and $\tau= \tau_0 + \int dt/R(t)$ denotes the conformal time.
Since $\sigma_0/f_{\rm PQ} \simeq (g/\lambda)^{1/2} (\Phi_0/f_{\rm PQ}) \gg 1$, the oscillation of $\sigma$ will initially take place around $\sigma =0$. \\

In both the cases, i.e. case-I and case-II, described above, the first phase of the oscillation around the origin terminates when the amplitude of $\sigma \propto {\rm R}^{-1}(t)$ drops below $f_{\rm PQ}$,
i.e. at
\begin{equation}
t_c = \!\!
\begin{cases}
t_0  \! \left( \! \tfrac{{\rm R}(t_c)}{{\rm R}(t_0)} \! \right)^{3/2} \! \simeq  {\cal H}_{\rm inf}^{-1} \! \left( \! \tfrac{\sigma_0}{f_{\rm PQ}} \! \right)^{3/2} \!\!, \mkern-18mu 
& (t_c \leq t_{\rm rh})
\\
t_{\rm rh}  \! \left( \! \tfrac{{\rm R}(t_0)}{{\rm R}(t_{\rm rh}) } \! \right)^{2} \!\!\! \left( \! \tfrac{{\rm R}(t_c)}{{\rm R}(t_0)} \! \right)^{2}
\!\!\! \simeq  \! {\cal H}_{\rm inf}^{-4/3} t_{\rm rh}^{-1/3} \! \left( \! \tfrac{\sigma_0}{f_{\rm PQ}} \! \right)^{2}  
\mkern-14mu
& (t_c > t_{\rm rh}) \, ,
\end{cases}
\end{equation}
with $t_{\rm rh}$
being the time of the reheating and $t_0$ being the time at the end of the slow roll inflation.
Here,
we assumed that the Universe is dominated by matter during the reheating epoch, i.e. ${\rm R}(t) \propto t^{2/3}$ for $t_0 <t<t_{\rm rh}$, and $t$ measures physical time.
At $t>t_c$, the amplitude of $\sigma$ is less than $f_{\rm PQ}$, and hence its oscillation takes place around $f_{\rm PQ}$.
If the PQ symmetry does not get restored during the oscillation phase either thermally or non-thermally, the energy density of the radial field will be dominated by the zero-mode, $\rho_\sigma(t \geq t_c) \simeq (\lambda f_{\rm PQ}^4/4) \, [{\rm R}(t_c)/{\rm R}(t)]^{3}$. 
Consequently at $t>t_c$, the oscillation of $\sigma$ can be treated as $\sigma$ particle with mass $m_\sigma = \sqrt{2 \lambda} f_{\rm PQ} $ setting at rest~\cite{Abbott:1982hn}, which dominantly decay into
ultra-relativistic
axions.

\section{Thermalisation and Thermal Production of Axions} \label{sec:th-prod} 

If the decay process of the $\phi$ field is sufficiently efficient, reheating the Universe to a high temperature, 
such a scenario can be envisaged in SM gauge invariant models of inflation~\cite{Allahverdi:2006iq,Allahverdi:2011aj},
axions may thermalise with the cosmic plasma. Later on they decouple from the plasma with thermal distributions.

Note that the axion field, $a \equiv  \theta f_a$, couples to the SM particles via $f_{\rm PQ}$-suppressed couplings. Nevertheless, such interaction can lead to the thermalisation of axions.
Before
the EW symmetry breaking, the axion interaction rate with the SM particles is dominated by its couplings to quarks $q$ and gluons $g$ via the axion coupling to gluons, $a \,  G^{b \mu \nu} \widetilde{G}_{\mu \nu}^b/f_a$~\cite{Graf:2010tv}~\footnote{In DFSZ models the axion interaction rate with the SM fields is dominated by the axion tree level coupling to quarks after EW symmetry breaking~\cite{Salvio:2013iaa}. However, the thermalisation of axions occurs at temperature much higher than the EW scale, and in such a case the dominant contribution to the interaction rate arises from the  axion-gluon coupling.}.
This is true for all axion models.
The relevant processes are then
\begin{flalign*}
&&  g + a \leftrightarrows q + \bar q,~ q + a   \leftrightarrows &  q + a,~ \bar q + a \leftrightarrows \bar q + a  \quad \quad \\
\text{and} 
&&  g + a   \leftrightarrows &  g + g,
\end{flalign*}
which give rise to the following interaction rate~\cite{Masso:2002np}~\footnote{The tree level thermally averaged interaction rate quoted above, $\Gamma_a n_{a, \rm eq}$ where $ n_{a, \rm eq}$ denotes the equilibrium number density of axions, is roughly the same as the one obtained using thermal field formalism~\cite{Salvio:2013iaa} for sufficiently small gauge coupling, $g_s$ which corresponds to sufficiently high temperature. They differ significantly when $ g_s \gtrsim 1$ which corresponds to $T \lesssim 5 \times 10^3$~GeV.
Since we are interested in axions with large $f_a$ which decouple from the plasma at $T \gg 5 \times 10^3$~GeV, the simple tree level calculation which gives the axion interaction rate, Eq.~(\ref{eq:Gamma_chi_thermal}), and also used in Appendix~\ref{sec:annihilation} is sufficient for our purpose. }
\begin{equation} \label{eq:Gamma_chi_thermal}
\Gamma_a \simeq 7 \times 10^{-6} \left(\frac{\alpha_s}{1/35}\right)^3 \frac{T^3}{f_a^2}
\, .
\end{equation}
Assuming that the SM quarks and gluons are part of the thermal bath,
the Boltzmann equation governing the number density of axions is given by~\cite{Kolb:1990vq}
\begin{equation} \label{eq:nchieom}
\dot n_a + 3 {\cal H} n_a =  \Gamma_a (n_{a, \rm eq} -  n_a )
\, ,
\end{equation}
where $ n_{a, \rm eq} = (\zeta(3)/\pi^2)T^3 $ denotes the equilibrium number density of axions with $\zeta$ being the Riemann zeta function,
and $ {\cal H} = (\pi^2 g_{*}/90 )^{1/2}   (T^2/M_{\rm P})$ is the Hubble expansion rate 
during radiation domination epoch with $g_{*}$ being the effective number of relativistic dof contributing to the total radiation energy density~\footnote{Here, we consider the evolution of $n_a$ during radiation domination, i.e for $T \leq T_{\rm rh}$. However, the Universe may have been exposed to temperature higher than $T_{\rm rh}$~\cite{Chung:1998rq,Mazumdar:2013gya}. Nevertheless from the axion production point of view, $T_{\rm rh}$ is effectively the maximum temperature as axions produced at $T>T_{\rm rh}$ gets diluted away by the entropy produced from $\phi$ decay~\cite{Salvio:2013iaa}.}.
Introducing
the function $\eta_a \equiv n_a/n_{a, \rm eq}$ and changing the dependence from time to $x \equiv T_{\rm rh}/T$,
where $T_{\rm rh}$ is the reheating temperature, Eq.~(\ref{eq:nchieom}) can be rewritten as
\begin{equation} \label{eq:etachieq}
x^2 \frac{d \eta_a}{d x} = K (1 - \eta_a) \, ,
\end{equation}
where
\begin{align}
K \equiv x \frac{\Gamma_a}{ {\cal H}} 
\simeq & \, 5\times 10^2 \left( \frac{g_*}{100}\right)^{-1/2} \!\! \left( \!\frac{\alpha_s}{1/35} \!\right)^3  \nonumber\\
& \!\! \times\left( \! \frac{T_{\rm rh}}{10^{10} {\rm GeV}} \! \right) \!\!
\left( \! \frac{f_{\rm PQ}/N_{\rm DW}}{10^{10} {\rm GeV}} \! \right)^{-2} .
\end{align}
Clearly, axions reach full thermal equilibrium with the SM particles if $K \gg 1$, i.e. when $T_{\rm rh}$ is sufficiently high.
Assuming that $g_{*}$ remains constant during the course of integration, Eq.~(\ref{eq:etachieq}) can be easily solvable~\cite{Masso:2002np}
\begin{equation} \label{eq:etachi}
\eta_a =  1- \ e^{K(x^{-1} - 1)} \, ,
\end{equation}
where we assumed that $ \eta_a(x=1)=0$.
Axions decouple from the plasma when $K \sim x$ (equivalently $\Gamma_a \sim \mathcal{H}$). In other words, axions decouple at
\begin{equation} \label{eq:Tchidec}
T_{a, \rm dec} \simeq 
10^7 \, {\rm GeV} \!
\left( \frac{g_*}{100}  \right)^{1/2} \!\!
\left( \! \frac{\alpha_s}{1/35} \! \right)^{-3} \!\!
\left( \! \frac{f_{\rm PQ}/N_{\rm DW}}{10^{10}~ {\rm GeV}} \!\right)^{2} .
\end{equation}
Since axions with $f_{\rm PQ} \gtrsim 4 \times 10^8$~GeV decouple at $T \gg m_Z$, where $m_Z$ denotes the $Z$-boson mass, they are  colder than photons at the time of last scattering as photons get reheated by the annihilation of other SM particles when the latter become non-relativistic. Therefore the contribution of thermally produced axions to $N_{\rm eff}$ is quite small~\cite{Weinberg:2013kea,Salvio:2013iaa}.
Nevertheless
as a consequence of the above discussion, the axions produced non-thermally at temperature $\ll T_{a, \rm dec}$ will never be in thermal contact with the plasma, and hence keep their abundance and momenta.
We elaborate on this in Appendix~\ref{sec:annihilation}.

\section{Non-thermal production of Axions} \label{sec:nonth-prod}

Let us now consider the decay of the coherent oscillation of the radial component of the PQ field, $\sigma$. If the PQ symmetry does not get restored during the initial phase of oscillation, which takes place around $\sigma = 0$, $\rho_\sigma$ will be dominated by the zero mode of $\sigma$.
For $t \geq t_c$, the oscillation of $\sigma$ continues around $\sigma = f_{\rm PQ}$ with initial amplitude $\sim f_{\rm PQ}$.
As the coherent oscillation of $\sigma$ behaves as $\sigma$ particles at rest, the latter cannot scatter into other species. However, they can decay into the particle species to which they couple. For instance, the $\sigma$ particles couples to axions via the vertex $ \tilde \sigma \partial_\mu  a \partial^\mu  a/f_{\rm PQ} $, where $\tilde \sigma = \sigma - f_{\rm PQ}$, allowing them to decay into axions with the following rate
\begin{equation} \label{eq:Gamma_chi_1}
\Gamma( \sigma \rightarrow  2 a) = \frac{1}{32 \pi} \frac{m_\sigma^3}{f_{\rm PQ}^2} = \frac{ \lambda^{3/2}}{8 \sqrt{2}  \pi} f_{\rm PQ}  \,.
\end{equation}

Moreover, the $\sigma$ particles can decay into species other than axions.
For example, in the KSVZ-like models~\cite{Kim:1979if} (see Appendix~\ref{sec:KSVZ} for a brief review), $\sigma$ couples to the 
extra heavy coloured fermions $Q$ via the vertex $(m_{Q}/f_{\rm PQ}) \, \sigma \, \bar Q Q $ leading to the following $\sigma$ decay rate
\begin{equation} \label{eq:Gamma_Q_1}
\Gamma(\sigma \rightarrow 2Q)
= \frac{3 m_Q^2 m_\sigma}{8 \pi f_{\rm PQ}^2} 
\left( 1 - \frac{4 m_Q{}^2}{m_\sigma^2}  \right) ^{3/2}
\, ,
\end{equation}
where for concreteness we assumed that the heavy quarks are colour triplets. The perturbative decay of $\sigma$ into extra heavy quarks is only allowed if $m_Q < m_\sigma/2$~\footnote{The decay of $\sigma$ to extra heavy quarks with $m_Q > m_\sigma/2$ can take place via non-perturbative effect during the fist phase of $\sigma$ oscillation as the extra heavy quarks become effectively massless during parts of each oscillation of $\sigma$~\cite{Greene:1998nh}. However once the amplitude of $\sigma$ drops below $f_{\rm PQ}$, the extra heavy quarks cannot be made massless and hence this non-perturbative decay channel is no more open.}.
Similarly, in the DFSZ-like models~\cite{Dine:1981rt} (see Appendix~\ref{sec:DFSZ} for a brief review), $\sigma$ couples to the two Higgs doublets,
\begin{equation} \label{eq:Gamma_H12_1}
\Gamma(\sigma \rightarrow 2 H_{1,2})
\simeq
\frac{\lambda_{S1,2}^2}{8 \pi m_\sigma} f_{\rm PQ}^2
\simeq \frac{\lambda_{S1,2}^2}{8 \pi \sqrt{2 \lambda} } f_{\rm PQ}
\, ,
\end{equation}
where $\lambda_{S1,2}$ have to be $< (v_{\rm EW}/f_{\rm PQ})^2$ with $v_{\rm EW}$ being the EW scale, in order not to affect the EW symmetry breaking~\cite{Volkas:1988cm}. Comparing Eqs.~(\ref{eq:Gamma_Q_1}) and (\ref{eq:Gamma_H12_1}) to Eq.~(\ref{eq:Gamma_chi_1}), we can see that $\sigma$ will decay mostly into axions in KSVZ-like models provided that $m_Q> m_\sigma/2$ or $m_Q< m_\sigma/(2 \sqrt{3})$, and similarly in DFSZ-like models as long as $\lambda >(v_{\rm EW}/f_{\rm PQ})^2$. 
As a result, 
entirely or at least large portion of the energy stored in the $\sigma$ field will ultimately be transferred to axion field.

We now proceed to estimate the axion energy density produced from the decay of the radial component of the PQ field. For simplicity, we will assume that the Universe is dominated by matter during the reheating epoch~\footnote{In principle, the effective equation of state during the reheating phase can be different from that of a matter dominated Universe, i.e. the equation of state parameter, $\omega_{\rm eff} >0$~\cite{Podolsky:2005bw}, in which case the energy density of the $\sigma$ field will be less diluted due to the slower expansion rate during the reheating phase, $ {\rm R}(t) \propto t^{2/[3(1+\omega_{\rm eff})]}$. As a result the abundance of the extra-relativistic axions due to $\sigma$ decay will be larger leading to a more stringent bound on the axion parameter space. For a review on reheating, see~\cite{Allahverdi:2010xz}.}. 
We further assume that all the $\sigma$ particles  instantaneously decay at $t=t_{\rm d}$ which can take place during inflaton domination (i.e. $t_{\rm d} < t_{\rm rh}$) if $ \lambda \geq (256 \pi^4 g_*/45)^{1/3} (T_{\rm rh}^2/f_{\rm PQ} M_{\rm P})^{2/3}$. Otherwise, the $\sigma$ decay process occurs during the radiation domination epoch (i.e. $t_{\rm d} > t_{\rm rh}$).

The energy density of $\sigma$ at $t=t_{\rm c}$ is roughly $(\lambda/4) f_{\rm PQ}^4$. Later at $t=t_{\rm d}$, $\rho_\sigma $ becomes $(\lambda/4) f_{\rm PQ}^4 [{\rm R}(t_{\rm c})/{\rm R}(t_{\rm d})]^3$. Assuming a sudden transition from inflaton domination to radiation domination at $t=t_{\rm rh}$, the energy density stored in $\sigma$ particles can be expressed as 
\begin{eqnarray} \label{eq:rhosigma0}
\rho_\sigma(t_{\rm d}) \simeq \frac{\lambda}{4} f_{\rm PQ}^4
\begin{cases}
\left( \frac{t_{\rm c}}{t_{\rm d}} \right)^{2}
&  (t_{\rm d} \leq t_{\rm rh})
\\
\left( \frac{t_{\rm c}}{t_{\rm rh}} \right)^{2}
\left( \frac{t_{\rm rh}}{t_{\rm d}} \right)^{3/2} 
&  (t_{\rm d} > t_{\rm rh} > t_c)
\\ \left( \frac{t_c}{t_{\rm d}} \right)^{3/2}
&  (t_{\rm d} >  t_c > t_{\rm rh})
\, ,
\end{cases}
\nonumber
\\
\end{eqnarray}
where
$t_c \simeq \mathcal{H}_{\rm inf}^{-1} (\sigma_0/f_{\rm PQ})^{3/2}
\leq t_{\rm rh}$ 
if $\sigma_0/f_{\rm PQ} \leq (90/\pi^2 g_*)^{1/3} (\mathcal{H}_{\rm inf} M_{\rm P}/T_{\rm rh}^2)^{2/3}$; otherwise, $t_c \simeq t_{\rm rh}(\mathcal{H}_{\rm inf}t_{\rm rh})^{-4/3} (\sigma_0/f_{\rm PQ})^{2}$.
The energy density of $\sigma$ comes to dominate the energy density of the Universe before it decays if
the following condition is violated:
\begin{equation}
\lambda <
0.15\times \frac{{\cal H}_{\rm inf}^8 M_{\rm P}^{10}}{g_*(T_{\rm rh})\sigma_0^{12} T_{\rm rh}^4 f_{\rm PQ}^2}
\, .
\end{equation}
The number density of $\sigma$ particles at $t_{\rm d}$ is $n_\sigma (t_{\rm d})= \rho_\sigma(t_{\rm d})/m_\sigma$.
These $\sigma$ particles decay dominantly into axions, and hence the number and energy density of axions at $t_{\rm d}$ are $n_a(t_{\rm d}) \simeq 2n_\sigma (t_{\rm d})$ and
\begin{eqnarray}
\rho_a(t_{\rm d}) \!
& \simeq & \!
2 [m_a^2 + (m_\sigma/2)^2]^{1/2} n_\sigma (t_{\rm d})
\nonumber \\
& \simeq & \!
[1 + (2 m_a/m_\sigma)^2 ]^{1/2} \rho_\sigma(t_{\rm d})
\nonumber \, ,
\end{eqnarray}
respectively~\footnote{Here we assume a monochromatic momentum distribution for axions due to the instantaneous decay of $\sigma$ particles at $t = t_{\rm d}$. In general, the decay of $\sigma$ particle takes place over an extended period of time leading to a smeared momentum distribution for axions due to the expansion effect~\cite{Kaplinghat:2005sy}.}.
In the range of interest, $f_{\rm PQ} \gtrsim 10^{8}$~GeV, axions are relativistic at the era of photon decoupling  ($z_{\rm dec} \simeq 1090$~\cite{Ade:2015xua}) and hence the factor $2 m_a/m_\sigma$ can be safely ignored. Thus with the help of Eq.~(\ref{eq:rhosigma0}), the axion energy density at the time interval $\max(t_{\rm d},t_{\rm rh}) < t < t_{\rm eq}$, where $t_{\rm eq}$ is the time at matter-radiation equality, is given by
\begin{equation} \label{eq:rhochi}
\rho_a (t)  \simeq  
\frac{\lambda}{4} f_{\rm PQ}^4
\begin{cases}
\left( \frac{t_{\rm c}}{t_{\rm d}} \right)^{2}
\left( \frac{t_{\rm d}}{t_{\rm rh}} \right)^{8/3}
\left( \frac{t_{\rm rh}}{t} \right)^{2}  
&  (t_{\rm d} \leq t_{\rm rh})
\\
\left( \frac{t_{\rm c}}{t_{\rm rh}} \right)^{2}
\left( \frac{t_{\rm rh}}{t_{\rm d}} \right)^{3/2}
\left( \frac{t_{\rm d}}{t}\right)^{2} 
&  (t_{\rm d} > t_{\rm rh} > t_c)
\\ \left( \frac{t_c}{t_{\rm d}} \right)^{3/2}
\left( \frac{t_{\rm d}}{t}\right)^{2} 
&  (t_{\rm d} >  t_c > t_{\rm rh})
\, .
\end{cases}
\end{equation}
If the PQ radial field does not come to dominate the energy density of the Universe before it decay, Eq.~(\ref{eq:rhochi}) can be rewritten as
\begin{align}
\frac{\rho_a}{\rho_{\gamma}} \simeq & \,
0.37  \left( \! \frac{g_*(T_{\rm rh})}{100} \! \right)^{1/3} \!\!
\left( \! \frac{10^{13}\,\mathrm{GeV}}{\mathcal{H}_{\rm inf}} \! \right)^2 \nonumber\\
& \times\left( \! \frac{\sigma_0}{M_{\rm P}} \! \right)^{3}  \!\!
\left( \! \frac{f_{\rm PQ}}{10^{15}\,\mathrm{GeV}} \! \right)^{1/3} \!\!
\left( \! \frac{T_{\rm rh}}{10^{10}\,\mathrm{GeV}} \! \right)^{4/3}
\label{eq:axion_abundance1} 
\end{align}
for $t_{\rm d} \leq t_{\rm rh}$, and
\begin{align}
\frac{\rho_a}{\rho_{\gamma}} \simeq & \,
0.04 \left( \! \frac{g_*(T_{\rm rh})}{100} \! \right)^{1/4} \!\!
\left( \! \frac{\lambda}{10^{-11}} \! \right)^{1/4} \!\!
\left( \! \frac{10^{13}\,\mathrm{GeV}}{\mathcal{H}_{\rm inf}} \! \right)^2  \nonumber\\
& \times\left( \! \frac{\sigma_0}{M_{\rm P}} \! \right)^{3}  \!\!
\left( \! \frac{f_{\rm PQ}}{10^{15}\,\mathrm{GeV}} \! \right)^{1/2} \!\!
\left( \! \frac{T_{\rm rh}}{10^{10}\,\mathrm{GeV}} \! \right)
\label{eq:axion_abundance2} 
\end{align}
for $t_{\rm d} > t_{\rm rh}$,
where $\rho_\gamma = (\pi^2/15)T^4$ denotes the energy density of photons.

\section{Constraints on Axion Parameter Space} \label{sec:constraints}

The axion parameter space is subjected to a plethora of experimental, astrophysical and cosmological bounds. We first review the most stringent ones
and then discuss the bound arising from the decay of the coherent oscillation of $\sigma$ into axions.

\begin{itemize}
\item{Supernovae:
Considering an extra energy loss channel in stars due to the emission of axions and comparing this to observations enables one to set upper bounds on the axion couplings and hence lower bounds on the axion decay constant, $f_a= f_{\rm PQ}/N_{\rm DW}$. The most stringent and model independent bound on $f_a$ arises for the obsevation of the supernova SN1987A signal, where the axion emission due to the nucleon bremsstrahlung ${\rm N} \, {\rm N} \rightarrow {\rm N} \, {\rm N} \, a$, if present, would have shortened neutrino burst duration (for a review, see e.g.~\cite{Raffelt:1996wa} and references therein). This places the following bound on $f_a$~\cite{Raffelt:2006cw}
\begin{equation} \label{eq:SN1987A}
f_a/\tilde{C}_{\rm N} \gtrsim 2 \times 10^9~{\rm GeV}
\, ,
\end{equation}
where $\tilde{C}_{\rm N} = ({\rm Y}_p C_p^2 + {\rm Y}_n C_n^2)^{1/2}$,
$C_p$ and $C_n$ are axion-nucleon couplings (${\cal L}_{a \rm NN} = C_{\rm N} \partial_\mu a \bar {\rm N}  \gamma^\mu \gamma_5 {\rm N}/2 f_a$ for $N=p,n$), and ${\rm Y}_p=0.3$ and ${\rm Y}_n=0.7$ are the proton and neutron fractions, respectively.
In the KSVZ models $C_p=-0.47(3)$ and $C_n=-0.02(3)$, whereas in the DFSZ models $C_p=-0.617+0.435 \sin^2 \beta \pm 0.025$ and $C_n=0.254-0.414 \sin^2 \beta \pm 0.025$ with $\tan\beta$ being the ratio of VEVs of two
Higgs doublets~\cite{diCortona:2015ldu}. Substituting for these values in Eq.~(\ref{eq:SN1987A}), we have 
\begin{equation}
f_a \gtrsim (2\textendash4) \times 10^8 \, {\rm GeV}
\, .
\end{equation}
Note that the neutrino burst duration of supernova SN1987A is less sensitive to the axion-nucleon coupling for 
$f_a \lesssim 6 \times 10^5 \, {\rm GeV}$ \cite{Burrows:1988ah,Raffelt:1996wa}, since axions with smaller $f_a$ would have been trapped at eariler stages. Nevertheless, their interaction with oxygen nuclei could have induced excitations in the oxygen nuclei resulting in the release of gamma ray that would have been seen at the Kamiokande detector~\cite{Engel:1990zd}.
As a result axions with $f_a \lesssim 2 \times 10^5 \, {\rm GeV}$ are ruled out.}

\item{Globular clusters:
Another bound on $f_a$ arises from the observation of the globular clusters~\cite{Raffelt:1989xu}. The possible axion energy loss via
Primakof process would accelerate the helium consumption reducing
the helium-burning lifetimes of the horizontal-branch stars. This places an upper bound on the axion-photon coupling (${\cal L}_{a \gamma \gamma} = g_{a \gamma \gamma} \, a F^{\mu \nu} F_{\mu \nu}/4$),  $g_{a \gamma \gamma} < 6.6\times10^{-11}\,\mathrm{GeV}^{-1}$~\cite{Ayala:2014pea}, where $g_{a \gamma \gamma} = \alpha_{\rm em}/(2 \pi f_a) \, [E/N-1.92(4)]$ with
$\alpha_{\rm em}$ and $E/N$ being the fine structure constant and the ratio of the electromagnetic to colour anomaly, respectively~\cite{Kaplan:1985dv}.
This translates to a lower bound of roughly $3 \times 10^7\,\mathrm{GeV}$ and $1 \times 10^7\,\mathrm{GeV}$ on $f_a$ for 
KSVZ and DFSZ models, respectively. 
The same argument of axion energy loss in the core of globular clusters stars can be used to constrain the axion-electron coupling relevant for the DFSZ models.
The axion-electron coupling would lead to the emission of axions from the core of red giants in globular clusters via the bremsstrahlung process $e+Ze \rightarrow e+Ze+a$. The observation of red giants places an upper bound on the axion-electron coupling [${\cal L}_{a e e} = g_{a e e} \, a \bar e \gamma_5 e$ with $g_{a e e}=m_e \cos^2\beta/(3 f_a)$], $g_{a e e}  \lesssim 4.3 \times 10^{-13}$~\cite{Viaux:2013lha}. This translates to a lower bound on the axion decay constant, $f_a \gtrsim 4.0 \times 10^8 \, \cos^2 \beta\,\mathrm{GeV}$.
 }

\item{White dwarfs:
Moreover, the axion-electron coupling
 $g_{a e e}$  
can also be constrained from the observation of white dwarfs. If $g_{a e e}$ is large it would increase the cooling rate of white dwarfs due to axion emission, which places an upper bound of $3 \times 10^{-13}$ on $g_{a e e}$~\cite{Bertolami:2014wua}. This translates to a lower bound of $6 \times 10^8 \, \mathrm{GeV}\,\cos^2 \beta$ on the axion decay constant.}

\item{Laboratory and hot dark matter bounds:
In addition to the astrophysical bounds discussed above, laboratory experiments rule out axions with $f_a \lesssim \mathcal{O}(10\textendash 10^2)\,\mathrm{GeV}$~\cite{Cheng:1987gp}. Thus in short, astrophysical observations and laboratory experiments rule out axions with decay constant $f_a \lesssim (2\textendash4) \times 10^8\,\mathrm{GeV}$, except for a possible small window, $ 2 \times 10^5 \, {\rm GeV} \lesssim f_a \lesssim 6 \times 10^5 \, {\rm GeV}$, particular to KSVZ-like models. We note that axions with $f_a$ in this window are ruled out from cosmological considerations. Axions can be produced thermally if $T_{\rm rh}>T_{a,\rm dec}$, where $T_{a,\rm dec}$ is given by Eq.~(\ref{eq:Tchidec}). 
In particular, axions with decay constant $f_a \lesssim \mathcal{O}(10^7)\,\mathrm{GeV}$ can be produced thermally and decouple from the plasma after the QCD phase transition. Hence, they contribute to the radiation density (not necessarily as an effectively massless dof) and later on act as hot DM, which sets an upper bound of around 1~eV on the mass of axion or equivalently a lower bound of around $6 \times 10^6$~GeV on $f_a$~\cite{Archidiacono:2013cha}. This rules out KSVZ-axions with $f_a$ in the small window not ruled out by astrophysical observations.
}

\item{Dark matter abundance:
On the other hand, the axion decay constant can be bounded from above. 
The first upper bound on $f_a$ arises from the requirement that the abundance of cold axions today does not
exceed the observed DM abundance, which implies that~\cite{Turner:1985si}
\begin{equation}
f_a \lesssim 7 \times 10^{11} \langle \theta_i^2 \rangle^{-0.84} \, {\rm GeV}
\, .
\end{equation}
Typically, $\langle \theta_i^2 \rangle^{1/2}$ is ${\cal O}(1)$, and in such a case, $f_a \lesssim 7 \times 10^{11}  \, {\rm GeV}$. However, in principle $\langle \theta_i^2 \rangle^{1/2}$ can be smaller than $\mathcal{O}(1)$, relaxing the upper bound on $f_a$.
}

\item{Isocurvature bound: if the PQ symmetry is broken before or during the early stages of inflation, large quantum fluctuations along the massless angular direction, $\delta \theta= N_{\rm DW} {\cal H}_{\rm inf}/(2 \pi f_{\rm PQ, eff})$ develop~\footnote{Similarly, the radial field can acquire quantum fluctuations of ${\cal O}({\cal H}_{\rm inf})$ if it is sufficiently light during inflation, $m_\sigma \ll {\cal H}_{\rm inf}$. This may contribute to the isocurvature perturbations as the radial field dominantly decays into axions.}. For sufficiently large PQ scale, axions do not thermalise with the cosmic plasma [see Eq.~(\ref{eq:Tchidec2})], and hence the fluctuations along the angular direction show up on the CMB sky as isocurvature perturbations with the following power spectrum~\cite{Harigaya:2015hha}
\begin{equation}
{\cal P}_{\rm Sc}= \frac{4}{\langle \theta_i^2 \rangle}
\left( \frac{N_{\rm DW} {\cal H}_{\rm inf}}{2 \pi f_{\rm PQ,eff}} \right)^2
\left( \frac{\Omega_a}{\Omega_{\rm CDM}} \right)^2
\, ,
\, 
\end{equation}
where $\Omega_a h^2=  0.2 \langle \theta_i^2 \rangle (f_a/10^{12}~{\rm GeV})^{1.19}$ is the abundance of CDM axions~\cite{Turner:1985si} and $h$ is the scaled Hubble parameter. 
The recent measurement of CMB~\cite{Ade:2015xua} constrains the CDM abundance, $\Omega_{\rm CDM} h^2= 0.1198 \pm  0.0030$ (at 95\% C.L.), 
and the CDM
uncorrelated isocurvature perturbations, $\alpha_c \equiv {\cal P}_{\rm Sc}/({\cal P}_{\rm Sc} + {\cal P}_\zeta) < 0.003$ where ${\cal P}_\zeta= 2.206^{+0.155}_{-0.145} \times 10^{-9}$ (at 95\% C.L.) is the amplitude of adiabatic perturbations. This places the following upper bound on the axion decay constant
\begin{equation}
f_a < 9.84 \times 10^7 \,{\rm GeV} \! \left(\!\! \frac{f_{\rm PQ, eff}}{\langle \theta_i^2 \rangle^{1/2} N_{\rm DW} {\cal H}_{\rm inf} } \! \right)^{0.84}
\!\! . \mkern-24mu
\end{equation}
The above bound need not be applied if $\phi$ is a moduli field.
}

\item{Superradiance:
Another interesting upper bound on $f_a$ arises from the consideration of the supperradiance effect of astrophysical rotating black holes~\cite{Penrose:1969pc}. Axion with large decay constant has a compton wavelength comparable to the size of the astrophysical black holes thus forming a bound system with different energy levels~\cite{Zouros:1979iw,Arvanitaki:2009fg,Arvanitaki:2014wva}. Such axions can then superradiate extracting rotational energy and angular momentum from the black hole through consecutive scatterings off the ergosphere and hence populating several energy levels. Axions can then emit gravitational waves via different processes resulting in a continuous extraction of angular momentum from the host black hole~\cite{Arvanitaki:2009fg,Arvanitaki:2014wva}. This would result in the absence of highly spinning black holes in a mass range corresponding to the range of $f_a$ for axions involved in superradiance. The measurement of the spin of stellar mass black holes disfavours axions with decay constant in the range $3 \times 10^{17} \, {\rm GeV} \lesssim f_a \lesssim 10^{19} \, {\rm GeV}$~\cite{Arvanitaki:2014wva}.}

\item{Dark radiation:
Axions are massless at the classical level, as they are protected by a shift symmetry, 
but they acquire a small mass, $m_a = 5.70(6)(4) \, {\rm eV} (f_a/10^{6}{\rm GeV})^{-1}$~\cite{diCortona:2015ldu},
due to the QCD instanton effect.
As a result axion is a natural candidate for dark radiation provided that $f_a$ is sufficiently large. 
Hence, they contribute to the effective number of relativistic dof other than photons $N_{\rm eff}$, which is defined via the relation that parametrises the total radiation  density of the Universe
\begin{equation}
\rho_{\rm rad} = \rho_\gamma \left[ 1 + \frac{7}{8} \left( \frac{T_\nu}{T_\gamma} \right)^4 N_{\rm eff} \right]
\, .
\end{equation}
Here, 
the neutrino-to-photon temperature
ratio is given by $T_\nu/T_\gamma=(4/11)^{1/3}$ with the assumption of exactly three neutrino flavors.
The extra contribution from relativistic axions can be estimated as
\begin{equation} \label{eq:deltaNeff}
\Delta N_{\rm eff} = N_{\rm eff} - N_{\rm eff}^{\nu} =  \frac{8}{7} \left( \frac{11}{4} \right)^{4/3} \frac{\rho_a}{\rho_\gamma}
\, ,
\end{equation}
where $N_{\rm eff}^{\nu}$ is the contribution from three active neutrinos.
Note that Eq.~(\ref{eq:deltaNeff}) applies only for axion species that remains relativistic till the era of photon decoupling.

The observation of CMB sets an upper bound $ N_{\rm eff} < 3.13 \pm 0.64$ with $95\%$ CL
at the time of photon decoupling~\cite{Ade:2015xua}.
In the standard cosmological model, $N_{\rm eff}^{\nu}$ is slightly larger than 3 neutrino species ($N_{\rm eff}^{\nu} = 3.046$) due to partial reheating of neutrinos when electron-positron pairs annihilate transferring their entropy to photons~\cite{Mangano:2005cc}. Thus, $\Delta N_{\rm eff} \leq 0.724$ at $95\%$ CL, which puts an upper bound on the abundance of relativistic axions. This can be used to constrain the parameter space of axion models.
}
\end{itemize}

Now we turn to the case where axions are produced non-thermally from the decay of the coherent oscillation of the radial component of the PQ field, $\sigma$.
Such axions have very large initial momenta and, furthermore, they can be very light provided that $f_a$ is sufficiently large.  
It is important to note here that for sufficiently large $f_a$ axions are never in thermal contact with the plasma and hence keep their initial abundance and momenta, see Eq.~(\ref{eq:Tchidec}).
Therefore, such axions are most likely to act as dark radiation contributing to $N_{\rm eff}$.
From Eqs.~(\ref{eq:axion_abundance1}), (\ref{eq:axion_abundance2}) and (\ref{eq:deltaNeff}), one can see that for large $f_a$,
$\Delta N_{\rm eff}$ can be much larger than $0.724$. In other words, too much axion radiations are produced. This further constrains the axion parameter space.

\begin{figure}[ht!] 
\centering
\includegraphics[width=8.5cm]{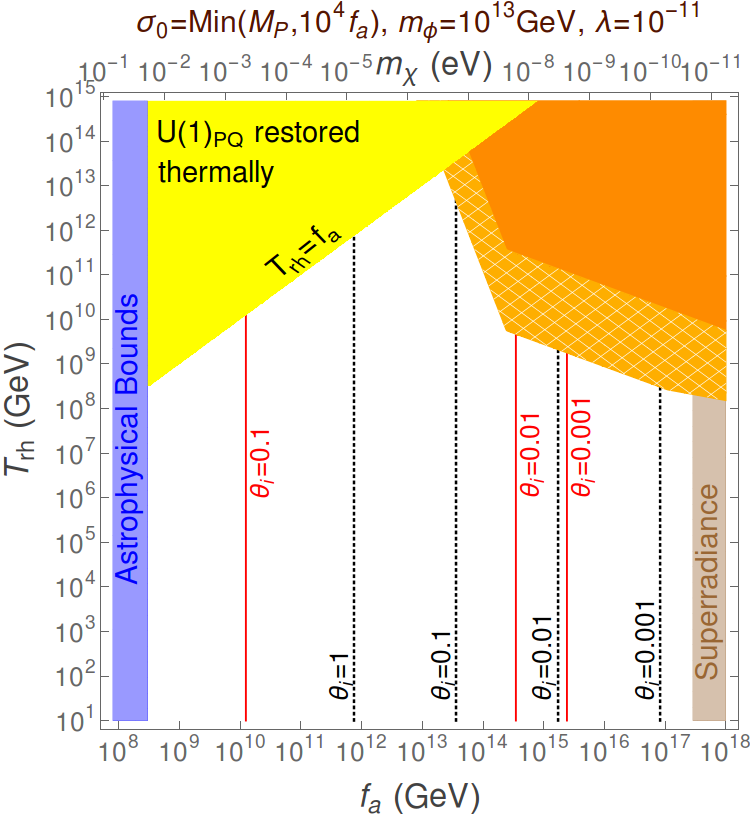}
\caption{Observational constraints on the reheating temperature $T_{\rm rh}$ and the axion decay constant $f_a$.
In the region shaded in dark orange, $\sigma$ comes to dominate the energy density of the Universe before it decays and hence is ruled out. The adjacent hatched region shaded in lighter orange is ruled out by the CMB bound on $N_{\rm eff}$. The blue region is ruled out by laboratory experiments and the observation of supernova SN1987A and globular cluster stars. In the yellow region, the PQ symmetry gets thermally restored.  The vertical dotted black lines indicate the CDM upper bound on $f_a$ whereas the vertical solid red lines refer to the isocurvature upper bound on $f_a$ for different values of misalignment angle. For $\theta_i=1$, the entire parameter space is ruled out by the CMB bound on the isocurvature perturbations. Note that the isocurvature bound {\it need} not apply if $\phi$ were a moduli field, the issue is rather model dependent.}
\label{fig:fig1}
\end{figure}

\begin{figure}[ht!] 
\centering
\includegraphics[width=8.5cm]{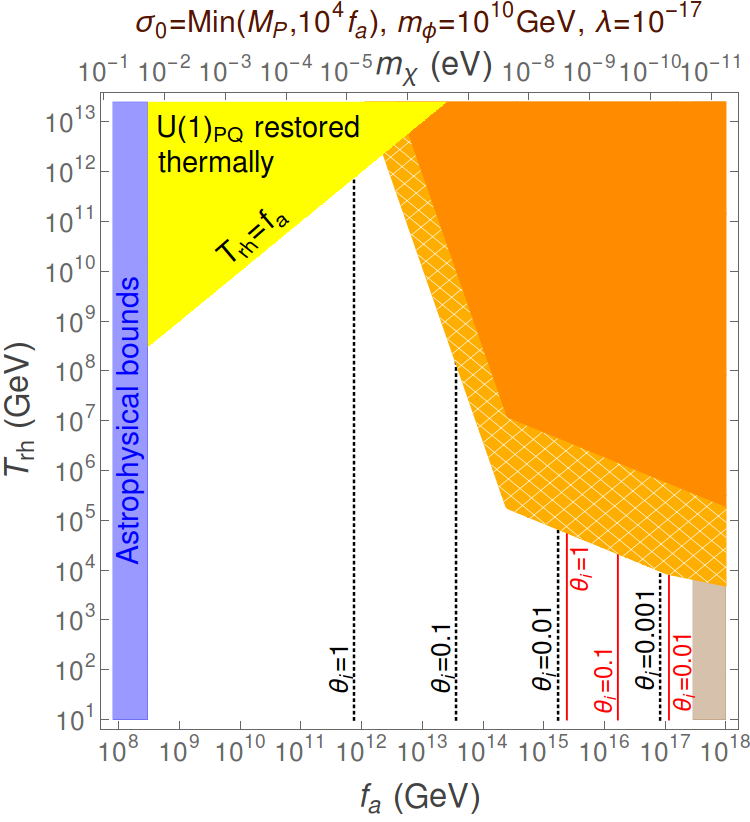}
\caption{Same as Fig.~(\ref{fig:fig1}) but for $m_\phi=10^{10}$~GeV and  $\lambda=10^{-17}$.}
\label{fig:fig2}
\end{figure}

We plot various observational constraints in Figs.~\ref{fig:fig1} and~\ref{fig:fig2} 
 where we fix the number of domain walls, $N_{\rm DW}=1$ and consequently
\begin{align}
f_a=f_{\rm PQ}. \nonumber
\end{align}
We have seen that different astrophysical observations rule out axions with $f_a \lesssim (2\textendash4) \times 10^{8}\,\mathrm{GeV}$ and $f_a \gtrsim 3 \times 10^{17}\,\mathrm{GeV}$. These bounds are shown in Figs.~\ref{fig:fig1} and~\ref{fig:fig2} by the blue and brown regions, respectively.
The current observational result on the DM abundance~\cite{Ade:2015xua} puts an upper bound on the axion decay constant, $f_a \lesssim 7 \times 10^{11} \langle \theta_i^2 \rangle^{-0.84} \, {\rm GeV}$.
In Figs.~\ref{fig:fig1} and~\ref{fig:fig2}, we  show  the DM bound on $f_a$ for different values of $\theta_i \simeq \langle \theta_i^2 \rangle^{1/2}$ 
by the dotted black lines. The regions to the right of these lines are ruled out.
%
%
%
%

On the other hand, if the initial amplitude of the oscillation satisfies $\sigma_0 \gtrsim 10^4 f_{\rm PQ}$,
the PQ symmetry can get restored non-thermally, which leads to the formation of topological defects~\cite{Kawasaki:2013iha}.
In order to avoid this, we require that
\begin{equation}
\sigma_0 = \min (M_{\rm P}, 10^4 f_{\rm PQ})
\, .
\end{equation}
It is worth noting here that the PQ symmetry is not restored during inflation as long as $f_{\rm PQ,eff}=\sigma_0 \gg {\cal H}_{\rm inf}/2 \pi$. However, the PQ symmetry can get restored thermally if $T_{\rm rh} \gtrsim f_{\rm PQ}= N_{\rm DW} f_a$.
Such a parameter space is shown in Figs.~\ref{fig:fig1} and \ref{fig:fig2} by the yellow regions. In principle, the Universe may have been exposed to temperatures much higher than $T_{\rm rh} $ during the reheating epoch~\cite{Chung:1998rq,Mazumdar:2013gya}. This makes the bound even more stringent.
We stress here that the restoration of the PQ symmetry and the subsequent formation of topological defects are only dangerous if $N_{\rm DW}>1$.
For models with
$N_{\rm DW}=1$, however, these topological defects are unstable and decay into cold axions
rendering the DM bound even stronger, $f_a \lesssim (4.6\textendash 7.2) \times 10^{10} \,\mathrm{GeV}$~\cite{Hiramatsu:2012gg,Kawasaki:2014sqa}.

For the case of $\phi$ being the inflaton,
we show the isocurvature bound on $f_a$ for $\theta_i=1, \, 0.1, \, 0.01 \, {\rm and} \, 0.001$ in Figs.~\ref{fig:fig1} and~\ref{fig:fig2}
by the solid red lines.
For $m_\phi = 10^{13}$~GeV (Fig.~\ref{fig:fig1}), the isocurvature bound is much stronger than the DM one. In this case the entire parameter space is ruled out when $\theta_i=1$.
For larger values of $N_{\rm DW}$, the isocurvature bound is even stronger.

The coherent oscillations of the radial part of the PQ field leads to the excitation of ultra-relativistic axions, which act as dark radiation and hence affect the expansion rate of the Universe. In a region of the parameter space, the radial field can come to dominate the energy density of the Universe, 
leading to axionic dark radiation dominated Universe. This scenario
does not produce the Universe
we live in and hence be ruled out. We show the region of the parameter space where this scenario occurs in dark orange shade in Figs.~\ref{fig:fig1} and \ref{fig:fig2}.
Even if the radial field does not come to dominate the energy density before it decays, the resultant axionic contribution to $N_{\rm eff}$ can exceed the CMB bound~\cite{Ade:2015xua}. This additionally rules out the hatched orange 
region shown in Fig.~\ref{fig:fig1} and \ref{fig:fig2}.
The constraint puts an upper bound on $T_{\rm rh}$, since if $T_{\rm rh}$ is large the $\phi$ field decays faster and 
the energy density of oscillating $\sigma$ field becomes relatively large.
With the help of Eqs.~(\ref{eq:axion_abundance2}) and (\ref{eq:deltaNeff}), the bound on the reheating temperature can be expressed as
\begin{align}
T_{\rm rh} \lesssim & \, 3.7 \times 10^{10} \, {\rm GeV} 
 \left( \! \frac{g_*(T_{\rm rh})}{100} \! \right)^{-1/4} \!\!\!
\left( \!\! \frac{\lambda}{10^{-11}} \!\! \right)^{-1/4} \!\!\!
\left( \! \frac{\sigma_0}{M_{\rm P}} \! \right)^{-3}
\nonumber\\
&\times   
\left( \!\! \frac{\mathcal{H}_{\rm inf}}{10^{13}\,{\rm GeV}} \!\!\right)^{2}
\!\!
\left( \!\! \frac{f_{\rm PQ}}{10^{15}\,\mathrm{GeV}} \!\! \right)^{-1/2}
\, .
\label{eq:reheating_temperature_bound} 
\end{align}
We also note that the duration of the oscillation of the $\sigma$ field becomes long if the initial amplitude is large $\sigma_0\lesssim 10^4 f_{\rm PQ}$, which
enhances the constraint in the large $f_a$ region.
Furthermore, if $\mathcal{H}_{\rm inf}$ (or the energy density of $\phi$) becomes small,
the energy density of SM plasma is reduced, which makes the abundance of ultra-relativistic axions relatively large.
As a result, a stronger constraint is obtained for a smaller value of $m_{\phi}\approx \mathcal{H}_{\rm inf}$, as shown in Fig.~\ref{fig:fig2}.
In Figs.~\ref{fig:fig1} and~\ref{fig:fig2}, we fix $\lambda$ to the maximum allowed value $\lambda \lesssim {\cal H}_{\rm inf}^2 M_P^2/\vert S \vert_0^4$
given by Eq.~(\ref{eq:lambdamax}).
Note that, since the axion energy density is given by $\rho_a \propto f_{\rm PQ}^4= (N_{\rm DW} f_a)^4$ [see Eq.~(\ref{eq:rhochi})],
for $N_{\rm DW}>1$ our bound becomes even stronger. Furthermore, to obtain conservative bounds, we assumed that Universe is dominated by matter during the reheating phase, i.e. the effective equation of state parameter, $\omega_{\rm eff} =0$.

\section{Conclusion} \label{sec:conclusion}

The PQ mechanism presents an elegant solution to the strong CP problem, and the angular field, the axion, can be a good DM candidate due to its largely suppressed coupling to all SM particles. However, axions being very light can also act as dark radiation if they are produced with large momenta at sufficiently late times. We showed that this can happen if the radial part of the PQ field was displaced from $f_{\rm PQ}$
due to an initial condition or a direct coupling to the inflaton/moduli field. 
The perturbative decay which we have discussed here happens when the PQ field oscillates coherently, during which
most of the co-moving energy density stored in these coherent oscillations gets transferred into light axions. The energy density of axions which contribute to the radiation energy density is constrained from number of observations listed above.  The bound is mostly relevant to larger values of the axion decay constant.

Axion DM with a large decay constant is expected to be probed by future experimental studies such as CASPEr~\cite{Budker:2013hfa}.
Since it is impossible to realize such a large PQ scale in the post-inflationary PQ symmetry breaking scenario~\cite{Kawasaki:1995vt},
one should seriously consider the pre-inflationary PQ symmetry breaking scenario if axion DM were to be found in such experiments.
We have seen that the cosmological evolution of the PQ field is quite non-trivial in such a scenario, and
the overproduction of ultra-relativistic axions leads to an upper bound on the reheating temperature, which further constrains the thermal history of the Universe. Furthermore, one can also expect many $\phi$ fields to oscillate simultaneously, either arising from inflation~\cite{Liddle:1998jc}, or due to multi-moduli fields, whose effects can be discussed by following similar arguments developed in this paper.

\section*{ACKNOWLEDGEMENTS}

AM would like to thank Masahide Yamaguchi for extensive discussion.
The work of A.M. is supported in part by the Lancaster-Manchester-Sheffield Consortium for Fundamental Physics under STFC grant ST/L000520/1. SQ is supported by a scholarship from the King Abdulaziz University.

\appendix

\section{KSVZ-LIKE MODELS} \label{sec:KSVZ}
In the KSVZ models~\cite{Kim:1979if}, the PQ field $S$ couples to the extra heavy 
quarks $Q_j$ via the vertices
\begin{equation} \label{eq:Qyukawa}
{\cal L} \supset - h_{j} ( \bar Q_L^j S Q_R^j + h.c.)
\, ,
\end{equation}
where we assumed real diagonal Yukawa coupling matrix. Under U(1)$_{\rm PQ}$, $S$ and $Q_j$ transform as $Q_j \rightarrow \exp(i \gamma_5 \alpha) Q_j$ and $S \rightarrow \exp(-2 i \alpha) S$, respectively, and all the other fields are invariant. The PQ symmetry gets spontaneously broken by the VEV of $S$, $\langle S \rangle = f_{\rm PQ}/\sqrt{2}$, giving mass to the extra heavy quarks, $m_{Q_j}= h_j f_{\rm PQ}/\sqrt{2}$. Moreover, the following couplings of $\sigma$ to the heavy quarks arise,
\begin{equation} \label{eq:Qyukawa2}
{\cal L} \supset - m_{Q_j}\frac{\tilde \sigma}{f_{\rm PQ}}  \bar Q^j Q^j 
+ \frac12 \frac{\partial_\mu a}{f_{\rm PQ}} \bar Q^j \gamma^\mu \gamma_5 Q_j
\, ,
\end{equation}
where $\tilde \sigma = \sigma - f_{\rm PQ}$. If we assume that there is at least one heavy quark $Q$ with $m_{Q} < m_\sigma/2$, the decay rate of the $\sigma$ into the heavy quark sector is given by
\begin{equation} \label{eq:Gamma_Q}
\Gamma(\sigma \rightarrow 2Q) = \frac{3 m_{Q}^2 m_\sigma}{8 \pi f_{\rm PQ}^2} 
\left( 1 -  \frac{4 m_{Q}^2}{m_\sigma^2}  \right) ^{3/2}
\, ,
\end{equation}
where $Q$ is the heaviest of the extra quarks with $m_Q < m_\sigma/2$. Equation~(\ref{eq:Gamma_Q}) has to be multiplied by a factor $N$ if we have instead $N$ nearly degenerate quarks. For concreteness, we assumed that $Q$ is colour triplet.

\section{DFSZ-LIKE MODELS}  \label{sec:DFSZ}

In the DFSZ models~\cite{Dine:1981rt}, $S$ couples to the two Higgs doublets which are charged under U(1)$_{\rm PQ}$ via the vertices
\begin{eqnarray} \label{eq:Higgss}
{\cal L} \supset \! & - &  \!
\vert S \vert ^2 (\lambda_{1S} \vert H_1 \vert ^2 + \lambda_{2S} \vert H_2 \vert ^2)
- \lambda_{S12} [S^2 H_1 \epsilon H_2 
\nonumber \\
& + & \!
 S^{*2} (H_1 \epsilon H_2)^*]
\, ,
\end{eqnarray}
where $\epsilon$ is the totally anti-symmetric matrix. The SM fermions also carry U(1)$_{\rm PQ}$ charges, but they do not couple to the $S$ field via renormalizable operators. After the PQ symmetry breaking the two Higgs doublets acquire extra mass terms $(\lambda_{1S} f_{\rm PQ}^2 /2) \vert H_1 \vert ^2$ and $(\lambda_{2S} f_{\rm PQ}^2 /2) \vert H_2 \vert ^2$, respectively. Here for simplicity, we assume that the mixing term is sufficiently small, $\lambda_{S12} \ll \lambda_{1S} , \lambda_{2S}$.
The EW VEV is then $v_{\rm EW} = \{ v_1^2 + v_2^2 -[ \lambda_{1S}/(2\lambda_{1}) + \lambda_{2S}/(2\lambda_{2})] \ f_{\rm PQ}^2 \}^{1/2}$ where $\lambda_{1,2}$ are respectively the quartic couplings of $H_{1,2}$. Since $f_{\rm PQ} \gg v_{\rm EW}$, the PQ field couplings to both the Higgs doublets have to be very small, $\lambda_{1S} , \lambda_{2S}, \lambda_{S12} < (v_{\rm EW}/f_{\rm PQ})^2$~\cite{Volkas:1988cm}.
The couplings of the radial excitation $\tilde{\sigma}=\sigma-f_{\rm PQ}$ to the Higgs fields are given by
\begin{eqnarray} \label{eq:Higgss2}
{\cal L} \supset  \!
& - & \!
\lambda_{1S} f_{\rm PQ} \tilde \sigma \vert H_1 \vert ^2 - \lambda_{2S} f_{\rm PQ} \tilde \sigma \vert H_2 \vert ^2
- \lambda_{S12}  f_{\rm PQ} \tilde \sigma  H_1 \epsilon H_2
\nonumber \\
& - & \!
\frac{\lambda_{1S}}{2} \tilde \sigma^2 \vert H_1 \vert ^2 - \frac{\lambda_{2S}}{2} \tilde \sigma^2 \vert H_2 \vert ^2 -
\frac{\lambda_{S12}}{2}  \tilde \sigma^2  H_1 \epsilon H_2 
\, .
\end{eqnarray}
The decay rates of $\sigma$ into these fields are respectively given by
\begin{equation} \label{eq:Gamma_H12}
\Gamma(\sigma \rightarrow 2 H_{1,2})
\simeq
\frac{\lambda_{S1,2}^2}{8 \pi m_\sigma} f_{\rm PQ}^2
\simeq \frac{\lambda_{S1,2}^2}{8 \pi \sqrt{2 \lambda} } f_{\rm PQ}
\, .
\end{equation}
There is also a cross coupling, which will lead to a similar decay rate of $\sigma$.

\section{ANNIHILATION OF NON-THERMALLY PRODUCED AXIONS} \label{sec:annihilation}

Let us consider the loss in the axion number density due to the scattering into SM particles. Ignoring Fermi blocking and stimulated emission, the Boltzmann equation governing the
time evolution of the axion number density can be written as
\begin{eqnarray} \label{eq:nchieq2}
\dot n_a + 3 H n_a  \! & = & \!  - \sum_{\rm spin} \!
\int \!
d \tilde{p}_a  d \tilde{p}_i
d \tilde{p}_1     d \tilde{p}_2
\
(2 \pi)^4 \times
\nonumber \\
&&  \!\!\!
\delta^{(4)} (P_a + P_i - P_1 -P_2)
\,
{\cal F}_{a} {\cal F}_{i, \rm eq} 
\
\vert {\cal M} \vert^2
\, ,
\nonumber \\
\end{eqnarray}
where $d \tilde{p}_j \equiv d^3p_j/[(2 \pi)^3 2 E_j]$. Here we ignore the axion production from the plasma. Again we assume that the SM particles ($i, \ 1 \  {\rm and} \ 2$) are in thermal equilibrium.
Using the definition of the cross section $\tilde \sigma$,
\begin{eqnarray} \label{eq:sigmachi}
\sum_{\rm spin} \!
\int && \!
d \tilde{p}_1
d \tilde{p}_2
(2 \pi)^4 \delta^{(4)} (P_a + P_i - P_1 -P_2)
\
\vert {\cal M} \vert^2 =
\nonumber \\
&&  \!
\tilde \sigma
v_{\rm Mol} \ 2 E_a 2 E_i
\, ,
\end{eqnarray}
Equation~(\ref{eq:sigmachi}) can be rewritten as~\cite{Gondolo:1990dk}
\begin{equation} \label{eq:nchieq3}
\dot n_a + 3 H n_a = -  \Gamma (a \, i \rightarrow 1 \, 2) \ n_{a}
\, ,
\end{equation}
with
\begin{eqnarray} \label{eq:gammachi2}
\Gamma (a \, i \rightarrow 1 \, 2)  = \frac{1}{n_{a}} \!
\int \!  d \tilde{p}_a   d \tilde{p}_i   \,
{\cal F}_{a} {\cal F}_{i, \rm eq} \,
\tilde \sigma
v_{\rm Mol} \, 2 E_a 2 E_i
\, \nonumber \\
\end{eqnarray}
being the averaged interaction rates where $v_{\rm Mol}=[(p_a^\mu p_{i \mu})^2 -m_a^2 m_i^2]^{1/2}/(E_a E_i)$ is the Moller velocity. In the relativistic limit, $ v_{\rm Mol} \, 2 E_a 2 E_i \simeq 2 s$ where $s = (m_a^2 + m_i^2) + 2 E_a E_i - 2 {\bf p}_a \cdot {\bf p}_i \simeq 2 E_a E_i (1 -  \cos \Theta_{a i}) $ is the squared total centre of mass (CM) energy, and $\Theta_{a i}$ is the angle between ${\bf p}_a$ and ${\bf p}_i$ ($\Theta_{a i} = \pi$ in the CM frame). We are free to evaluate Eq.~(\ref{eq:gammachi2}) in the CM frame.
Expressing ${\bf p}_a$ and ${\bf p}_i$ in polar coordinates, Eq.~(\ref{eq:gammachi2}) can be rewritten as
\begin{eqnarray} \label{eq:gammachi3}
 \Gamma (a \, i \rightarrow 1 \, 2) \!
&= & \!
\frac{1}{4 \pi^4 n_{a}} \int_{-1}^{1} \! \frac12 d\cos \Theta \int_{0}^{\infty} \!\! d p_a p_a^2 \int_{0}^{\infty} \!\! d p_i p_i^2
\nonumber \\
&& \!
\times {\cal F}_a {\cal F}_{i, \rm eq} \tilde \sigma_{\rm CM}
\, ,
\end{eqnarray}
where $\tilde \sigma_{\rm CM}$ is the cross section in the CM frame with no average over the internal dof.
For non-thermal axions produced from the decay of $\sigma$ particles, one can approximate the phase space distribution of axions as
\begin{equation} \label{eq:Fchi}
{\cal F}_a (p_a, t) = 2 \pi^2 n_a(t_{\rm d}) \left( \frac{{\rm R}(t_{\rm d})}{{\rm R}(t)} \right)^3
\frac{\delta(p_a-\frac{{\rm R}(t_{\rm d})}{{\rm R}(t)} p_{a}(t_{\rm d}))}{p_a^2}
\, ,
\end{equation}
where we assumed a sudden decay of $\sigma$ at $t=t_{\rm d}$. Again we focus on the axion
interactions with the SM quarks and gluons via the axion anomalous coupling~\cite{Graf:2010tv,Masso:2002np}:
\begin{enumerate}
\item $ g + a \leftrightarrows q + \bar q $,
\item $ q + a \leftrightarrows q + a $ and $ \bar q + a \leftrightarrows \bar q + a $,
\item $ g + a \leftrightarrows g + g $,
\end{enumerate}
which dominate the axion interaction rate with the SM particles at temperatures above the EW symmetry breaking scale. These
interactions lead to cross sections of the form $\tilde \sigma_{\rm CM} = A \ln (s/m_D^2) + B $~\cite{Masso:2002np}, where $m_D= \sqrt{8 \pi \alpha_s} \, T$ is the Debye mass, and  $A$ and $B$ are constants whose values respectively are
\begin{enumerate}
\item $A=0$  ,  $B= \frac{N_f}{6 \pi^2} \frac{\alpha_s^3}{f_{a}^2}$,
\item $A=\frac{N_f}{\pi^2} \frac{\alpha_s^3}{f_{a}^2}$, $B = - \frac{3 N_f}{4 \pi^2} \frac{\alpha_s^3}{f_{a}^2}$,
\item $A=\frac{15}{2 \pi^2} \frac{\alpha_s^3}{f_{a}^2}$  ,  $B = - \frac{55}{8 \pi^2} \frac{\alpha_s^3}{f_{a}^2}$,
\end{enumerate}
with $N_f=6$. Substituting $\tilde \sigma_{\rm CM}$ with the corresponding values of $A$ and $B$ for each of the above interactions into Eq.~(\ref{eq:gammachi3}) and summing up all the contributions, we obtain
\begin{eqnarray} \label{eq:gammachi4}
\Gamma (a \, i \rightarrow 1 \, 2) \! &=& \!
\{33 [\ln (\tilde p_a/T ) - \ln (2 \pi  \alpha_s) +\zeta'(3)/\zeta(3)
-\gamma]
\nonumber \\
&& \!
+6 \ln (2) - 29 \}
\frac{\zeta (3)}{2 \pi ^4} \frac{\alpha_s^3 T^3}{f_a^2}
\nonumber \\
&\simeq& \!
\{ 9.4 +  4.7 \ln(\tilde p_a/T) \} \times 10^{-6}
\left(\frac{\alpha_s}{1/35}\right)^3 \frac{T^3}{f_{a}^2}
\, ,
\nonumber\\
\end{eqnarray}
where $\tilde p_a = p_{a}(t_{\rm d}) ({\rm R}(t_{\rm d})/{\rm R}(t))$ and $\gamma$ is the Euler's constant. Clearly the factor $\ln(\tilde p_a/T)$ is constant 
since $T \propto {\rm R}^{-1}(t)$. As one would expect, $\Gamma (a \, i \rightarrow 1 \, 2)$ is slightly larger than $\Gamma (1 \, 2 \rightarrow a \, i)$ due to the monochromatic momentum distribution of axions produced from $\sigma$ decay. Moreover, the factor $\ln(\tilde p_a/T)$ slightly enhances $\Gamma (a \, i \rightarrow 1 \, 2)$ further if $\tilde p_a \gg T$. Again introducing the function $\eta_a= n_a/n_{a, \rm eq}$ and the independent variable $x=T_{\rm d}/T$, where $T_{\rm d}$ is the temperature corresponding to the time $t_{\rm d}$, the Boltzmann Eq.~(\ref{eq:nchieq3}) can be rewritten as
\begin{equation} \label{eq:etachi3}
x^2 \frac{d \eta_a}{d x} = - K  \eta_a \, ,
\end{equation}
where $K \equiv x  \Gamma(a \, i \rightarrow 1 \, 2) /{\cal H} $. Eq.~(\ref{eq:etachi3}) admits the following solution
\begin{equation}
\eta_a (x) =  \eta(x_{\rm d}) \ e^{K(x^{-1} -1)} \, .
\end{equation}
Since $x^{-1}=T/T_{\rm d} \leq 1$, axions keep their initial abundance if $K \ll 1$. In other words, axions produced non-thermally at
\begin{equation} \label{eq:Tchidec2}
T_{\rm d} \ll 10^7 ~ {\rm GeV} \left(\frac{\alpha_s}{1/35}\right)^{-3}
\left( \frac{g_*}{100}  \right)^{1/2}
\left( \frac{f_{\rm PQ}/N_{\rm DW}}{10^{10}~ {\rm GeV}}\right)^{2} ,
\end{equation}
will keep their initial abundance and momenta.



\end{document}